**Efficient Interfacial Solar Steam Generator with Controlled Macromorphology Derived from Flour via "Dough Figurine" Technology**

Zhengtong Li[1], Chengbing Wang[1,*], Zeyu Li[1], Lin Deng[1], Jinbu Su[1], Jing Shi[2], Meng An[2,*]

[1]School of Materials Science and Engineering, Shaanxi Key Laboratory of Green Preparation and Functionalization for Inorganic Material, Shaanxi University of Science and Technology, Xi'an, Shaanxi 710021, China

[2]College of Mechanical and Electrical Engineering, Shaanxi University of Science and Technology, Xi'an 710021, China

* To whom correspondence should be addressed.

E-mail: wangchengbing@gmail.com (C. Wang), anmeng@sust.edu.cn (M. An)

**Abstract**

Solar-driven interface steam generator (SISG) is a most promising technology for seawater desalination and wastewater purification. A shape- and size-controlled, low-cost, eco-friendly solar-absorber material is urgently desired for practical application of SISG. Herein, we proposed a facile, sustainable and scalable approach to produce tailored SISG with controlled macromorphology derived from flour via "dough figurine" technology which is originated from the China Han Dynasty. Three kinds of self-floated flour-based absorbers i.e. near-cylindrical (integrated), near-spherical (loose packing) and powdery (dense packing) absorber used as SISGs were discussed, we found that the macromorphology significantly influences water transport and interfacial thermal management of SISG, the integrated absorber has an overwhelming advantage, which possesses a high evaporation efficiency 71.9% at normal solar illumination. The proposed "dough figurine" technology breaks the limitations of the inherent geometry of reported biomass based SISG, which provides an important guidance for SISG use in remote and impoverished areas.

# 1. Introduction

The discovery of solar-driven interfacial steam generation (SISG) with the ultrahigh performance evaporation has sparked a wave of vigor and excitement due to its critical role in ubiquitous related applications, such as seawater desalination,[1, 2] power generation[3-5] and steam sterilization[6] and so on. Indeed, SISG overcomes many inherent obstacles in conventional solar-driven steam generation, i.e. the efficiency of the solar energy receiver, thermal management capabilities, high production costs and complex preparation processes.[7-9] Recently, various artificial materials of SISG equipment have been designed based on different novel mechanisms, such as plasmonic metal particles absorber,[10-12] carbon-based materials absorber,[13-15] and heat confinement layers[16, 17]. For example, in such plasmonic absorber, the plasmonic noble metal particles are deposited on the alumina template (AAO) or wood blocks, where the plasmon effect of plasmon particle and light traps induced by porous substrate (e.g. AAO, wood blocks) are utilized to enhance the absorption of solar energy.[10, 11, 18, 19] Such porous substrates can transport bulk water to the photo-thermal conversion interface by capillary force as well as provide the channel of steam overflow. Meantime, their low thermal conductivity effectively can reduce the heat conduction loss. In analogy, large amounts of porous substrates combined with traditional synthetic carbon materials are also developed as efficient solar distillation equipment, such as graphene,[20] reduced graphene oxide,[21, 22] graphene alkyne,[23] carbon nanotubes,[24] carbon power.[25] However, previous SISG materials with high efficiency steam evaporation often entail complex fabrication processes, high cost and difficult to recycle, which limits their application domain. Therefore, the challenge of developing a cost-effective, robust, environmentally friendly, SISG material remains to obtain a high photo-thermal conversion efficiency under ambient solar illumination.

Recently, biomass materials with natural ubiquitous micro- and macrostructures, such as mushroom,[26] lotus [27], cotton,[28] wood,[29] daikon,[30] louts leaf,[31] are developed as efficient solar steam generation device by a series of simple processing i.e. cutting, freeze drying and high-temperature carbonization (**Figure S1**). Some simple treatments for biomass materials have produced unparalleled advantages for synthetic materials (**Table S1 and S2**).[32] For example, mushroom with an umbrella-shaped structure (i.e. positive solid cone geometric structure) can possess a high evaporation efficiency 78% under 1 sun irradiation.[26] Natural wood is constructed as a high performance of SISG system. The evaporation efficiency of the carbonization poplar wood-based SISG device can reach 86.7% at 10 kW m$^{-2}$, which is a fascinating discover for the biomass materials application in SISG system.[33] Due to different natural mesoporous structures in natural wood, it is also used to investigate the effect of water transport and heat transfer on evaporation efficiency. The carbonized lotus seedpods with a high a evaporation efficiency (86.5% @1 SUN) mainly relied on their unique macroscopic cone shape and hierarchical mesopores and macropore structures.[27] Moldy bread, carbonized by a barbecue-like method, achieves 71.4% evaporation efficiency in a high humidity level (70%).[34] In our recent study, a kind of arched structure bamboo charcoal is developed as a high-performance absorber, which possess a larger area for absorbing sunlight and a smaller area for water absorption.[1] These studies of biomass materials-based SISG can provide a marvelous opportunity to lower cost and high evaporation efficiency. However, it is quite difficult to make a large-scale fabrication due to the inherent geometry structures of biomass materials. Therefore, a shape- and size-controlled, low-cost, eco-friendly biomass-based absorber is urgently desired for practical application of SISG

Flour, prepared by mechanical grinding of wheat, is the main raw material for cooking steamed bread and noodles. The production of wheat all over the world is very huge. Plenty of expired flour (> 10 million tons in China every year) is generated due to excessive production.[35-38] Therefore, it is desirable to explore novel value-added application. After two basic processes i.e. shaping and dehydration, four can be made as several typical macrostructures (**Figure S2**) with the large surface area beneficial for solar light absorption, as well as inner porous structures to achieve a better performance of water transport. After carbonization, the solar absorption capacity of flour-based absorber with near-columnar is greatly improved to 92.4% across the entire solar spectrum without affecting its mechanical properties. During preparation, only commercial deionized water without any expensive and toxic chemicals is added as a good binder/templating agent. Inspired by Chinese traditional craft, called "Dough Figurine (DF)" in the Han Dynasty, the flour mixing with water can form dough. Through several simple operations such as pinching, rubbing, pushing, etc., various macro geometries can be prepared. More importantly, the macroscopic morphology of the flour-based materials can preserve without any significant changes after the dehydration and carbonization. Therefore, these advantages of flour indicate its promising as a candidate material for solar absorber and searching for a high-performance evaporation structure. In this study, we utilized flour to create solar-driven interface steam conversion equipment.

Herein, inspired by the craft of DF, we prepared flour block with different shapes (near-columnar, cup-shaped, umbrella-shaped and near-spherical) which were freeze-dried and carbonized at a lower temperature (300 ℃). Porous carbon foam can be prepared easily, which satisfies all essential elements for interfacial steam generation system. In details, it can float freely on the water surface due to its lower mass density and the self-floating time can keep for a long time (more than 720 hours in seawater).

Meanwhile, it has high light absorbance as high as 92.4% from 200 nm to 2500 nm. The efficient water transport (carbonized flour with a rich pore structure) can be achieved by its rich pores and hydrophilic. We firstly discussed in detail the impact of different stacking forms of the same materials on optical absorption, water transport, and thermal management of SISG, we found that the macromorphology significantly influences the evaporation efficiency of SISG, the integrated absorber has an overwhelming advantage, which possesses a high evaporation efficiency 71.9% at normal solar illumination. The proposed "dough figurine" technology breaks the limitations of the inherent geometry of reported biomass based SISG, which provides an important guidance for SISG use in remote and impoverished areas.

## 2. Result and discussion

Similar with the common morphology of solar absorber in SISG system, the carbonized flour-based absorber with near-cylindrical (NC) geometry is fabricated and used to study the evaporation performance of SISG system.[10, 14, 15, 23, 39] **Figure 1** shows schematic diagram of making flour-based solar absorber. As shown in Figure 1(a) and 1(b), flour is a kind of biomass material obtained by mechanical grinding of wheat. The short growth cycle (one or two times a year) and large yield ($128.8*10^6$ t in China 2016)[40] of wheat show an overwhelming advantage over that of other biomass materials, which suggests a large-scale flour can be produced in a factory with low cost. In addition, plenty of flour can't keep for a long time due to some environmental factors, such as moisture, pests, etc. Herein, we utilized flour, an excellent candidate for biomass carbon sources, to create SISG equipment. Propelled by the Chinese traditional method called "DF" craft, flour-based absorbers with different geometries can be prepared (The details of preparation are shown in the experimental section). Figure 1(c) and 1(d) show macro model of the flour block sample with NC geometry before (FNC) and after (FFNC) freeze-drying process, where large amounts of nanopores form by adding and removing of water. It is obviously found that tight starch granules piled together on the surface of FFNC. The FFNC is carbonized under an air environment at 300 °C, which is more available compared with that of previously reported studies (Table S1). The easy access to prepare FFNC sample (low carbonization temperature and no gas protection during carbonization) is more conducive to large-scale fabrication of SISG equipment.

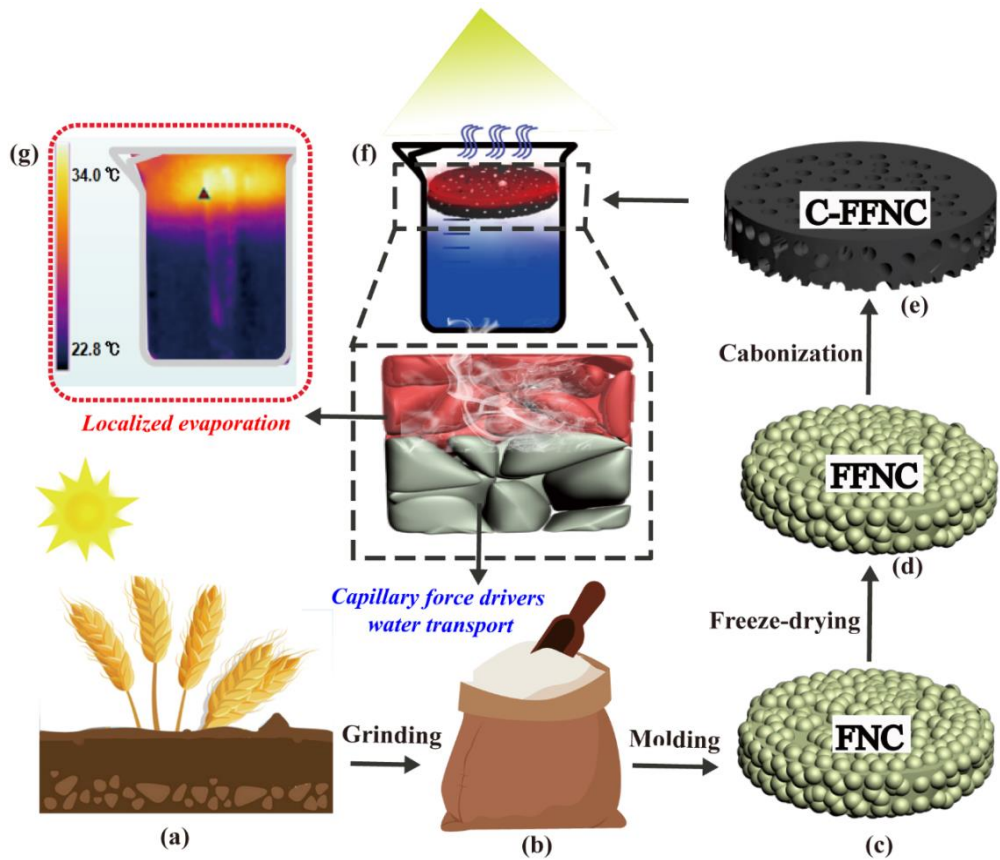

**Figure 1.** Schematic of preparation process of near cylindrical solar absorber based flour block. (a) Natural wheat. (b) Commercially available flour. (c) Preparation of flour block with near-cylindrical (FNC) structure using dough figurine technology. (d) Freeze-dried (FFNC) sample. (e) The carbonization process of freeze-dried (C-FFNC) sample. (f) The schematic picture of solar-driven interface steam generation (SISG)

device. The inset picture: the localized evaporation of C-FFNC can be enhanced. (g) The infrared photo of **SISG** device under one sun irradiation at 1min.

The carbonized freeze-drying flour block with near-cylinder structure, simplified as C-FFNC, is shown Figure 1e. During the process of fabrication, C-FFNC possesses a large mechanical strength (**Figure S3**) due to the absence of activator (such as $Na_2CO_3$, $K_2CO_3$, and KOH). Without plenty of other gas ($CO_2$ or $H_2$) formation, the inner wall of the tunnel is kept thick[35, 36] (Figure 2c). The C-FFNC sample as SISG sample can form localized evaporation (Figure 1f and inset picture), where water can be pumped spontaneously through its micro-scale channels under the synergistic assistance of capillary force and the absorptive nature of the hydrophilic starch and the vapor for fresh water can be generated under the nature solar light illumination.[27] As shown in Figure 1g, the infrared image of the SISG device under one sun radiation in one minute, the localized hot region can be obviously observed.

The microstructure plays a critical role on the water transport and solar absorption of the SISG system.[41] Figure 2a shows the outer surface of FFNC sample. It consists of starch granules with different sizes. After carbonization, multi-scale holes are formed on inner and outer surface of C-FFNC samples shown in Figure 2b and 2c. In addition, the porosity of samples is evaluated before and after carbonization using a mercury intrusion instrument. Based on the measured diameter of pores, the average values of the pore diameters are analyzed (**Figure S4**). The pores diameter of FFNC samples are mainly micro and mesopores (≤50nm) which are induced by dehydration in the freeze-drying process. In carbonization, the pore structures of C-FFNC samples collapses and reorganizes due to the burning of organics (Figure S5). Therefore, the pore structures with meso- and macrospores (≥50 nm) are dominated. The porosity is decreased from 80.3% to 59.8% while the corresponding average pore diameter is increased from 5.4 nm to 173.0 nm. The macro-pores surface of C-FFNC (shown in Figure 1c) provide the channel for the generated steam to escape. This prevents the generated steam from cooling in the original position, thereby reducing the evaporation efficiency. The inset structure of C-FFNC (Figure 1d) showed channels with different pore diameters, which can transport water from the bulk water to the hot region by capillary force. Further, these internal tunnel structures provide a rich heat exchange area that effectively reduces heat loss to the bulk water. The thermal conductivity of FFNC and C-FFNC are 0.12 W/m-K and 0.09 W/m-K (**Figure S6**), respectively.

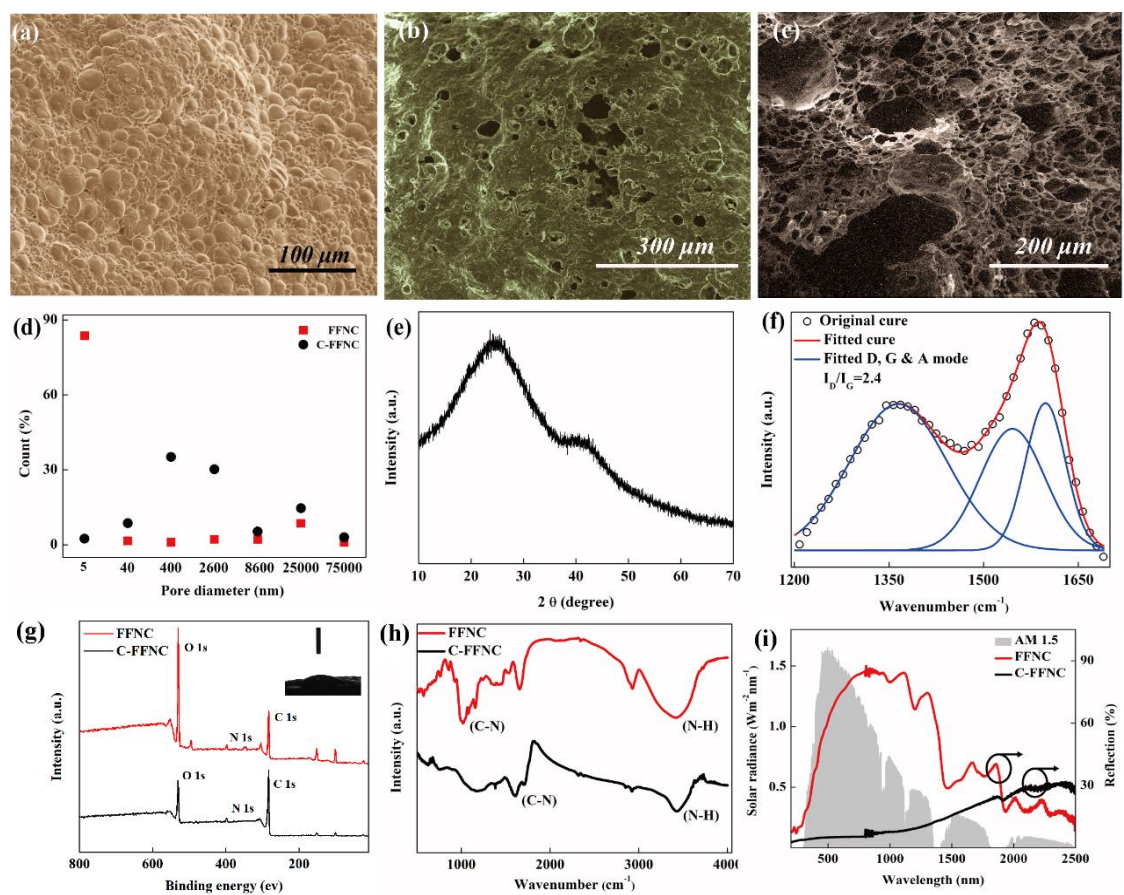

**Figure 2.** (a) SEM image of the surface of freeze-dried samples. (b) SEM image of the surface of C-FFNC sample. (c) SEM image of the inner structure of C-FFNC sample. (d) Porosity data statistics for FFNC sample and C-FFNC sample. (e) The XRD

diffraction pattern of C-FFNC. (f) Raman spectra of C-FFNC fitted by Gaussian functions. (g) The XPS spectra of FFNC and C-FFNC. Inset picture: contact angle test of C-FFNC. (h) The FTIR spectra of FFNC and C-FFNC. (i) Solar irradiance (AM 1.5 G) (blue, left side axis) and reflection (right side axis) of FFNC and C-FFNC.

The water transport of C-FFNC can be enhanced due to its hydrophilicity originating from its special components and carbonization condition. Flour, 17% protein and 25% carbohydrate as the main components, it is easily bonded by stirring after adding deionized water. Also, both carbohydrates and proteins have good hydrophilicity, so the formed dough also has good hydrophilicity which is conducive to the transfer of water from bulk water to air-liquid interface. The degree of carbonization is measured by X-ray diffractometry (XRD). As shown in **Figure 2e**, two broad diffractions were observed at the 2-theta of the degree of 21° and 40°, which suggests that the main structure in the C-FFNC sample is amorphous carbon. The stronger peak of 21° is a typical graphene reflection, indicating the presence of a majority graphene-like structure in the C-FFNC sample. And it can be proved by Raman spectra (Figure 2f). In the Raman spectra of C-FFNC, the D-band with disorder amorphous carbon and G-band with $sp^2$ vibration of graphite crystal can be observed at 1361 $cm^{-1}$ and 1560 $cm^{-1}$. In order to accurately analyze the data, we performed a sub-peak fitting of the measured Raman spectra, the $I_D/I_G$=2.4 what can prove that the C-FFNC sample have a higher degree of amorphous carbon.[27, 42, 43] In addition, we further explored surface chemical composition of C-FFNC by X-ray photoelectron spectroscopy (XPS) and Fourier transformed infrared spectroscopy (FTIR). The XPS spectrum of the sample before (FFNC) and after (C-FFNC) carbonization is shown in the **Figure 2g**, the main elements of the two including carbon, nitrogen, oxygen. The relative contents of C1s, N1s and O1s of FFNC are 55.39%, 2.49%, 42.15%, respectively. After carbonization, the relative content of N 1s of C-FFNC is increased to 2.67%, and corresponding to C-N/N-H. As the FTIR (Figure 2h) show, the C-N/N-H functional group can still be detected at the 1135cm$^{-1}$ and 3435cm$^{-1}$. The C-FFNC has excellent hydrophilicity and can be further confirmed by the contact angle test. The contact angle of its surface is about 34.6 °(inset picture of Figure 2g) which contribute its surface chemistry elements and its rough texture.

As it is mentioned above, C-FFNC can ensure that it floats on the water surface to transport a sufficient amount of water from the bulk water to the upper surface of the absorber due to its rich pore structures and hydrophilicity.[32] Besides, excellent optical absorption and effective heat management are critical to enhance evaporation efficiency. Therefore, it is very meaningful to further examine optical capabilities and thermal management. As shown in Figure 2i, the sample before carbonization, the FFNC has a very high reflection in the visible light band. After carbonization, the color of the sample turned black and the reflection was greatly reduced. Based on the formula:

$$\alpha = \frac{\int_0^\infty A(\lambda)[1-R(\lambda)]d\lambda}{\int_0^\infty A(\lambda)d\lambda}$$

where A(λ) is the solar radiant energy density at different wavelengths of AM 1.5 at atmospheric mass. R is the reflectance of the solar absorber at different wavelengths that is usually taken from 0.2 to 2.5μm. Therefore, the absorbance of C-FFNC sample is calculated about 92.4% and even it can reach 95.5% in the visible range (400-760 nm). As mentioned earlier, the stacking way of SISG devices is also an important topic that has to be considered (i.e. under the same conditions: whether an integrated material works well compared with the stacked multiple materials in a certain way.). So we carbonized samples of different shapes (i.e. near-cylindrical (integrated), near-spherical (loose packing) and powdery (dense packing)) were prepared to examine the placement effect of evaporation efficiency. Integrated samples, loosely packed samples, and densely packed samples are three common and important ways of stacking (Figure 4a). The C-FFNC is a sample of the monolithic material. Then we prepared some near spherical (NS) samples with a diameter of ~1 cm, and obtained loosely packed samples after freeze-drying, carbonization (i.e. called C-FFNS). Further, we pulverize some C-FFNS samples and pass the standard filtration to obtain a powder particle size of ~450μm as dense packed samples (i.e. called C-FFP). Unlike nanograde samples stacks, these three kinds of macro samples stacking are similar in terms of optical absorption, hydrophilicity and mechanical strength.[44, 45] We conclude that important factors affecting the evaporation efficiency of these three different stacking materials are their different thermal management. The mass changes of C-FFNC, C-FFNS and C-FFP have large differences (Figure 3a). To systematically evaluate the evaporation performance of C-FFNC, the mass change of water under normal solar illumination (1 sun) was recorded (Figure 3a). The evaporation rate and efficiency are 1.0 kg m$^{-2}$ h$^{-1}$ and 71.9%, respectively, which is much higher than that of pure water. Such high evaporation efficiency of C-FFNC mainly is explored based on the formula of evaporation rate, the net evaporation rate $\dot{m}$ can be expressed as

$$\dot{m}h_{fg} = A\alpha q_{solar} - A\varepsilon\sigma(T^4 - T^4_{environment})$$
$$-Ah(T - T_{environment}) - Aq_{water} \quad (1)$$

where $h_{fg}$ is the latent heat, A is surface area of the absorber facing the sun, $\alpha$ is the solar absorption, $q_{solar}$ is the solar flux, $\varepsilon$ is the emittance of the absorbing surface, $\sigma$ is the Stefan-Boltzmann constant (i.e., 5.67 × 10$^{-8}$ Wm$^{-2}$K$^{-4}$),[46] T is the surface temperature of the absorber, T$_{environment}$ is the temperature of the adjacent environment, h is the convection heat transfer coefficient, and $q_{water}$ is the heat flux to the underlying water, including conduction and radiation. The second and the third term on the right side of Eq. (1) denotes the heat loss to ambient by heat radiation and convection. According to Eq. (1), the evaporation rate can be enhanced by increasing heat energy absorption and decreasing heat loss (heat convection, heat radiation, and heat conduction).

For C-FFNC absorber, the solar light trap formed by the surface pores and the carbonized nature with black color can form a high optical absorption (95.5% in the visible range), which provide sufficient solar energy source. The thermal conductivity of flour-based carbon foam is 0.09 W/m-K, one order of magnitude lower than that of pure water (0.60 W/m-K at room temperature).[47, 48] The low thermal conductivity of C-FFNC enables the captured solar energy to be confined to the air-liquid surface, which reduces the heat conduction loss to the bulk water and promote the formation of localized hot area as shown in Figure 1(g). Moreover, the rich pore structures (porosity 59.79 %) and hydrophilicity of C-FFNC facilitates the water transport from bulk water transport to air-liquid interface. In addition, the previous studies suggested that the enthalpy of confined water is reduced compared with that of pure water. In C-FFNC absorber, the rich micro-scale pores of C-FFNC facilitates the reduced vaporization enthalpy, further enhancing evaporation rate.

Admittedly, precisely controlling the size and shape of SISG device, especially at a harsh environment, is challenging and subject to large uncertainties. Interestingly, our proposed flour-based absorber can be prepared into different geometries (shown in Figure S2), which promises to expand applications of SISG device and making existing applications more advantageous and cost-effective. Moreover, such advantage of flour-based absorber make flour as an optimal candidate material to explore the placement effect of absorber on evaporation rate. To systematically analyze the evaporation performance of flour-based absorber with stack ways, the mass change of water under normal solar illumination (1 sun) was recorded (Figure 3a). The mass change of C-FFNC, C-FFNS, and C-FFP are 0.54 kg/m², 0.41 kg/m² and 0.34 kg/m² in half an hour, respectively, which are 2.3, 1.7, and 1.5 times as high as that of pure water (Figure 3a). The corresponding evaporation rates are 1.0 kg m$^{-2}$ h$^{-1}$, 0.74 kg m$^{-2}$ h$^{-1}$ and 0.62 kg m$^{-2}$ h$^{-1}$, respectively. The evaporation efficiency ($\eta$) is defined as below:

$$\eta = \frac{h_{LV}\dot{m}}{C_{opt}q_i} \quad (2)$$

where $\dot{m}$ represents the evaporation rate (kg m$^{-2}$h$^{-1}$) after the subtraction of the evaporation rate without light illumination, $h_{LV}$ is the total enthalpy of the liquid-vapor phase change (sensible heat + phase-change enthalpy), $C_{opt}$ represents the optical concentration and $q_i$ represents the nominal solar illumination (1 kW m$^{-2}$).

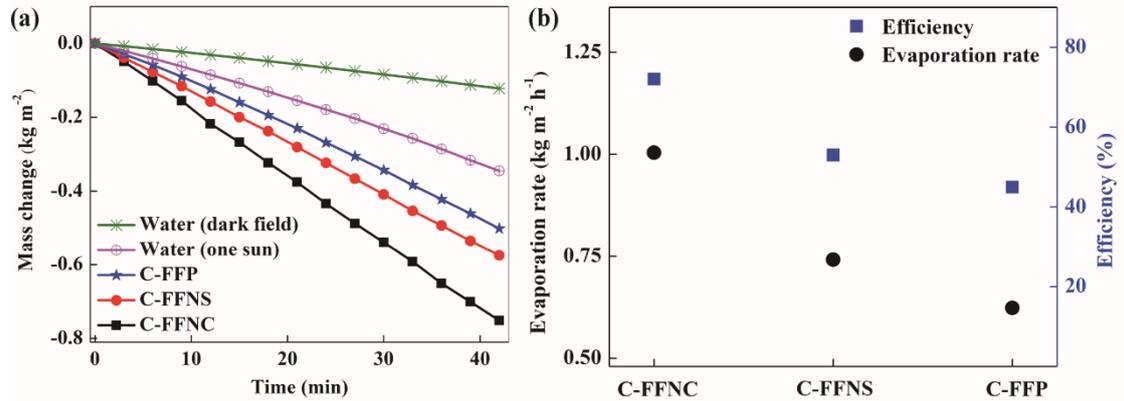

**Figure 3.** (a) Mass change of C-FFNC, C-FFNS, and C-FFP as well as water in the dark field and under 1 sun illumination. (b) The evaporation rate (left side axis) and evaporation efficiency (right side axis) of C-FFNC, C-FFNS and C-FFP.

The calculated energy efficiencies of evaporation for C-FFNC, C-FFNS, and C-FFP are presented in Figure 3b. It is obviously observed that the energy efficiency of C-FFNC can reach 71.9%, which are much higher than those of C-FFNS (~ 53.1%) and C-FFP (~ 44.7%). In other words, the C-FFNC absorber is the most effective for enhancing the energy efficiency among three kinds of samples. We will explore the potential mechanism for this trend from three main factors: heat localization, heat loss and water transport.

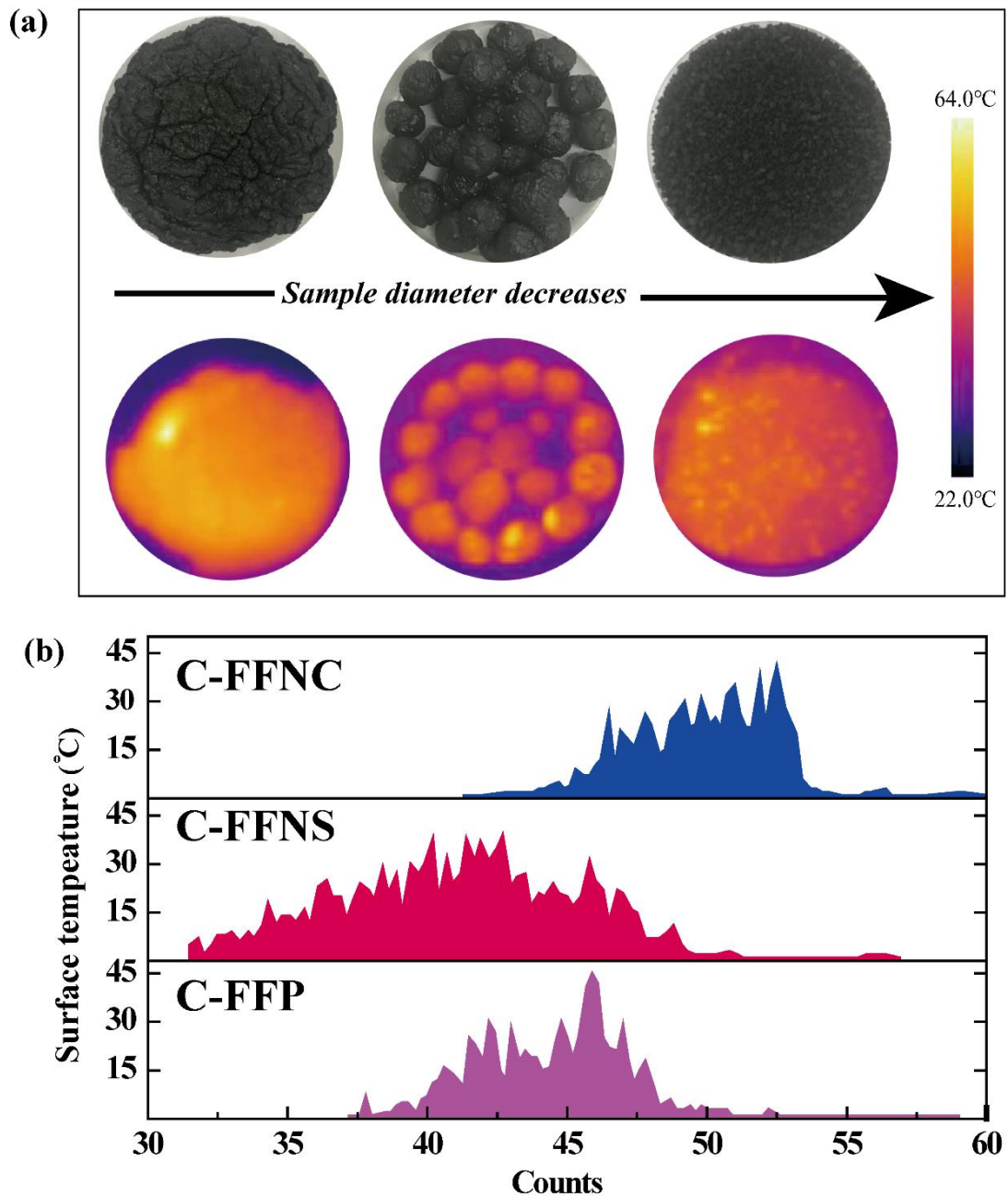

**Figure 4.** (a) The optical image and the infrared image of the C-FFNC/ C-FFNS/ C-FFP sample surface in 1 min under one sun irradiation. The radius of three samples are mainly about 5 cm, 1 cm, 425 μm for C-FFNC, C-FFNS and C-FFP, respectively. (b) Surface temperature statistical point distribution image of the (a) infrared images.

The heat transfer behavior of flour-based absorber with different sample diameters is systematically evaluated (Figure 4). The surface temperatures of C-FFNC, C-FFNS and C-FFP under 1 sun illumination are carefully measured via an IR camera. Figure 4 (a) plots the surface temperature image from IR camera, from which we calculated the temperature distribution of sample surfaces. Obviously, the average steady-state temperatures of the C-FFNC, C-FFNC, C-FFNS and C-FFP under 1 sun illumination reach 49.9 °C, 41.2 °C and 44.4 °C, respectively (Figure 4b). These surface temperature are much higher that of pure water (23 °C). Compared with thermal conductivity (0.09 W/m-K) of carbonized flour-based materials (the details are shown in Figure S4), the larger thermal conductivity of bulk water results in a considerable proportion of the absorbed energy for increasing the underlying water temperature instead of evaporation, which reduces the energy efficiency. More interestingly, the surface temperature of C-FFNC is higher than those of C-FFNS and C-FFP samples. Although the intrinsic porosity among samples with different diameters are close, such stacking with different samples diameters leads to this decreasing trend of water content inside carbonized floured-based absorber: C-FFP, C-FFNS, C-FFNC. The more water, the more the absorbed energy is used to heat underlying water. Therefore, the heat localization by for adiabatic material (C-FFNC), only the water on C-FFNC sample surface is heated and the underlying water will not cause large heat conduction loss. That is, almost all the heat is localized at the air-liquid interfacial, which results in the high surface temperature and fast evaporation.

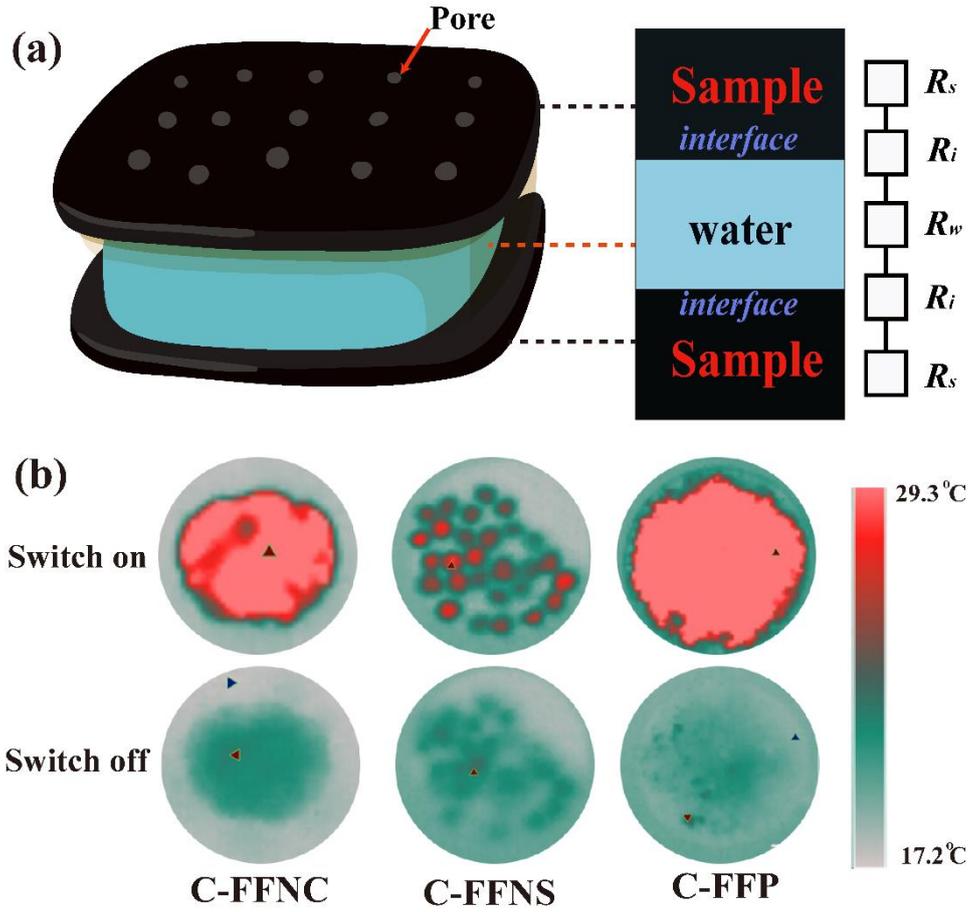

**Figure 5.** (a) The schematic diagram of thermal resistance between solid-liquid interfaces for samples floating on the bulk water. (b) The wetting process of C-FFNC, C-FFNS and C-FFP at 0s and 120s.

To further understand the high performance of evaporation of C-FFNC, the heat energy dissipated by interfacial thermal resistance is qualitatively analyzed.[49-51] The total thermal resistance of the floured-based absorber including flour-based materials and inside water is defined by introduce the interfacial thermal resistance $R_i$ shown in Figure 5(a),

$$R_t = R_w + R_s + R_i \qquad (3)$$

Here, $R_w$ and $R_s$ are the intrinsic thermal resistance of bulk water and flour-based sample. $R_i$ is the interfacial thermal resistance between water and flour-based sample, which is related to the interfacial properties of interfacial component materials and interface area. According to Eq. (3), among three samples (i.e., C-FFNC, C-FFNS and C-FFP), the interfacial properties are almost the same while the total interface area are obviously different due to different diameters of flour-based absorber. The diameters of C-FFNC, C-FFNS and C-FFP are about 5.0 cm, 1.2 cm, and $1.5 \times 10^{-4}$ cm, respectively. Specifically, the ratio of their interface areas is 1: 4.16: $3.32 \times 10^4$. Thus, the heat energy inside C-FFP absorber is severely dissipated by interfacial thermal resistance, leading to a low evaporation rate.

As shown in figure 5b, we explore the water transporting capability of samples with different stacking ways. C-FFNC, C-FFNS, and C-FFP samples are heated in oven and then are put into water directly under the lab environment (detail in supporting information). The cooling of the sample is mainly cooled by water nature evaporation. The faster the cooling rate, the faster the water transport capacity. Experiments show that water transport capacity decreases from C-FFP to C-FFNS to C-FFNC in turn.

As aforementioned, the evaporation efficiency (E. E.), thermal resistance (H. R.), water transportation (W. T.) relationships are:

$$E.E._{C\text{-}FFNC} > E.E._{C\text{-}FFNS} > E.E._{C\text{-}FFP.}$$

$$H.R._{C\text{-}FFNC} > H.R._{C\text{-}FFNS} > H.R._{C\text{-}FFP.}$$

$$W.T._{C\text{-}FFNC} < W.T._{C\text{-}FFNS} < W.T._{C\text{-}FFP.}$$

Since the three samples are of the same material, the wettability, porosity, and optical absorption capacity (Figure S7) are very close. Their difference in evaporation efficiency is mainly caused by the different heat resistance and water transport. The relationship between the speed trend of water transport (**Figure 5b, details in Figure S8**) and the evaporation efficiency is linearly negatively correlated, which further indicates that the thermal resistance is the main factor determining the different placement modes of the samples.

## 3. Conclusion

Inspired by the technique of dough figurine, we developed a new method for designing biomass-based SISG with controlled micromorphology derived from flour. The self-floated carbonized flour-based absorber with excellent mechanical properties, high efficiency of solar absorption, low thermal conductivity and rich pore structure, which fulfill all the require of SISG for efficient localized evaporation. This flour-based SISG with the common structure (near-cylindrical) can achieve a high evaporation efficiency about 71.93% at normal solar illumination. Furthermore, we first investigated the effects of different macroscopic accumulations of uniform absorber on evaporation efficiency, and explored in detail the effects of different cumulative modes of optical absorption, water transport, and thermal management. This study provides an important guidance for the large-scale use of SISG device in remote and impoverished areas.

## 4. Experiment Section

**The preparation process of flour dough with specific shape:**

The water content and mixing time play a critical role for ensuring the similar micro-structure of the samples. The flour (purchased from local supermarket) and deionized water is mixed in a ratio of 2:1. The time of mixing-process is about 15 minutes. According to the craft of DF, these samples with different geometric shapes (near-columnar, umbrella-shaped and near-spherical) were prepared (the details are shown in **Figure S2**) with hand.

**The preparation process of absorber material:**

Here, we take flour dough with near-columnar (FNC) as an example to introduce the process of carbonization. First, FNC sample is frozen at 0 ℃ in the refrigerator for 10 hours. Then, the frozen FNC sample is dehydrated in a freeze dryer for 24 hours. Finally, the freeze-dried sample (FFNC) is carbonized in a muffle furnace. The carbonization is conducted under the condition of heating to 300 ℃ for two hours at 5 ℃/min. The carbonized samples (C-FFNC) can be used as SISG device directly.

**Materials Characterization**:

The flour-based sample morphology was characterized via a high-resolution field-emission scanning electron microscope (SEM, FEI Verios 460 USA), Environmental scanning electron microscope (FEI Q45+EDAX Octane Prime), and the contact angle measurement through video optical contact angle measurement with measurement accuracy of 0.1 °. The characteristic spectral reflection peak of sample was mainly measured by an ultraviolet-visible-near-infrared spectrophotometer (Cary 5000). The sample surface temperature was measured via an IR camera (Fotric) with model LS5.5B4−0038 lens and FTIR spectra were obtained on a VERTEX 70 spectrometer. The porosity was measured by fully automatic mercury porosimeter (AutoPore IV 9510). The Phase testing was performed via an X-ray diffractometer (XRD, D8 advance, Bruker, Germany). The Raman spectra were measured via the micro-Raman spectrometer (Renishaw inVia Reflex, UK) with a CCD detector at room temperature. The elemental valence states were determined by X-ray photoelectron spectroscopy (XPS, ThermoFisher ESCALAB 250Xi, monochromatic Al Kα. Thermal Decomposition and Metamorphic Test Method Synchronous Thermal Analysis TG-DSC (TA, NETZSCH). Thermal conductivity of samples are measured using hot disk method (TPS 2500S), Standard sieve for powder filtration (US standard).

**Solar steam generation experiments.**

The C-FFNC was placed in a beaker rolled by commercial expanded polyethylene in order to reduce the ambient effect on experimental results. The salt water (3.5 wt% of salt, which is the same salt concentration as seawater) is placed in the beaker. The samples were illuminated under one-sun intensity with a xenon lamp (CEL-HXF300,

AM1.5 filter). The mass during the evaporation process was measured by an electron microbalance (AR224CN) with an accuracy of 0.0001 g and the data were transmitted to the personal computer (PC). The room temperature and humidity during the tests were controlled at 24 ℃ and 30%, respectively.

**Supporting Information.**

The Supporting Information is available free of charge on the ACS Publication Website.

**Notes**

The authors declare no competing financial interest.

**Acknowledgement**

This work is financially supported by the National Natural Science Foundation of China (Nos. 51562020 and 51575253) and Natural Science Research Start-up Fund of Shaanxi University of Science and Technology (2018GBJ-10). We are grateful to Guilong Peng, Xiaoxiang Yu for useful discussions.

# Table of Contents

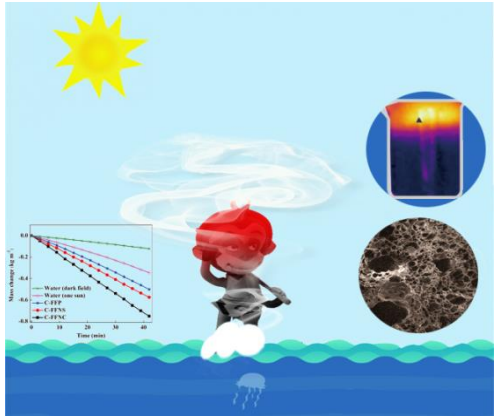

**Efficient Interfacial Solar Steam Generator with Controlled Macromorphology Derived from Flour via "Dough Figurine" Technology**

# Supporting information

**Efficient Interfacial Solar Steam Generator with Controlled Macromorphology Derived from Flour via "Dough Figurine" Technology**

Zhengtong Li[1], Chengbing Wang[1,*], Zeyu Li[1], Lin Deng[1], Jinbu Su[1], Jing Shi[2], Meng An[2,*]

[1]School of Materials Science and Engineering, Shaanxi Key Laboratory of Green Preparation and Functionalization for Inorganic Material, Shaanxi University of Science and Technology, Xi'an, Shaanxi 710021, China

[2]College of Mechanical and Electrical Engineering, Shaanxi University of Science and Technology, Xi'an 710021, China

* To whom correspondence should be addressed. E-mail: wangchengbing@gmail.com (C. Wang), anmeng@sust.edu.cn (M. An)

**Table S1** Comparison of important parameters of commercially available materials for use as SISG equipment

| Material | Carbonization temperature (℃) | Carbonization condition | Evaporation efficiency | Experiment temperature(℃) | Experiment humidity (%) | Ref. |
|---|---|---|---|---|---|---|
| mushrooms | 500 | Ar@12h | 78%@1sun | 28 | 41 | **1** |
| basswood | 500 | Hot plate | 57.3%@1sun | constant | constant | **2** |
| basswood | * | Flame treatment | 72%@1sun | 26 | 40 | **3** |
| Sponge | 500/700/900 | $N_2$/2h | *(2.5-fold) | 22.5 | * | **4** |
| wood | 500 | Hot plate/60s | 86.7%@10sun | * | * | **5** |
| wood | 500 ℃ | * | 74%@1sun | * | * | **6** |
| melamine foams | 400/550/ 700 | $N_2$/2h | 87.3%@1sun | 25 | 50 | **7** |
| daikon | 750 | $N_2$/2h | 85.9%@1sun | 28 | * | **8** |
| louts | 500 | $N_2$ | 86.5%@1sun | 28 | * | **9** |
| bread | 400 | Homemade furnace | 71.4%@1sun | 21 | 70 | **10** |

| flour | 300 | Air/2h | 71.4%@1sun | 24 | 30 | **Our work** |

Ps. *On behalf of the reference does not specifically mention the relevant conditions

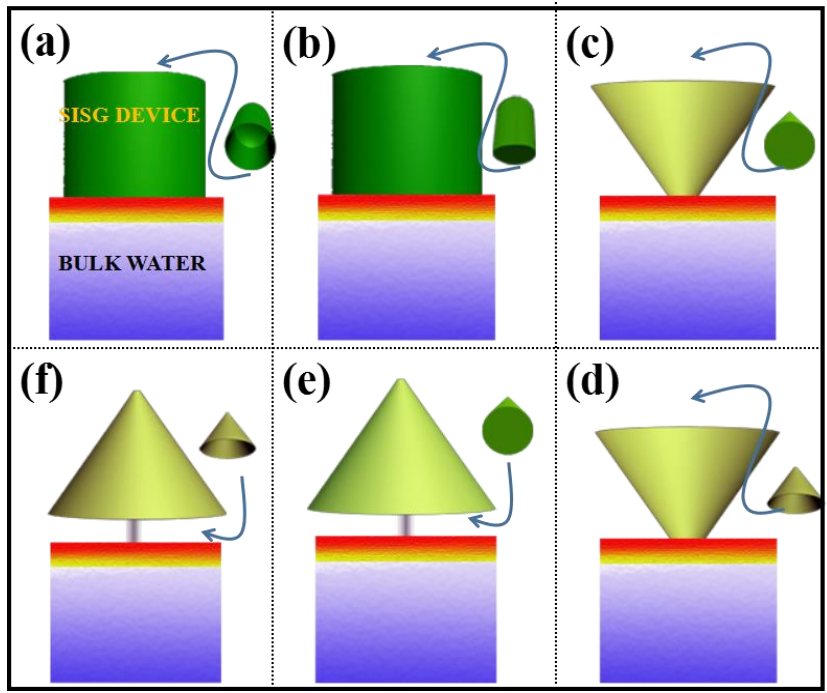

**Figure S1**. Several typical special structure macroscopic absorber models. The detail of mode is show in Table S2.

The models shown in Figure S1 are currently typical solar absorbers with special macro geometry. We summarize the corresponding geometric model structure (see Table S2) and its unique advantages. And we have made some structures using flour-based materials (see Figure S2 for details). This section mainly shows that the flour-based material is a solar absorber preparation material with great potential.

Table S2. Advantage Analysis of Several Typical Special Structure Macroscopic Absorber Models

| No. | Ref. | Style | Characteristics |
|---|---|---|---|
| a | **11, 12** | Hollow cylinder | Multiple reflection hit inner wall |
| b | **13** | Solid cylinder | long surface(minimizing the energy loss from the top surface and maximizing the energy gain from the side surfaces) |
| c | **9, 14** | Inverted solid cone | Large surface(light), small path(water) |
| d | **15,16** | Inverted hollow cone | Multiple reflection hit inner wall; small path(water) |
| e | **1** | Positive solid cone | large surface-projected area ratio(smaller temperature increase)---suppressed convection and radiation losses |
| f | **16, 17** | Positive hollow cone | large surface-projected area ratio |

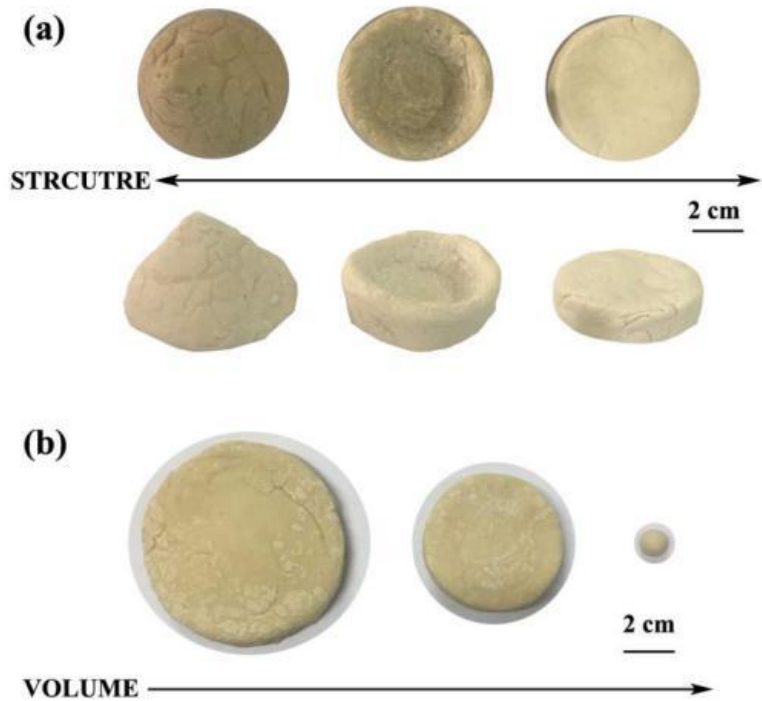

**Figure S2.** Flour block preparation of different shapes and volumes by dough figurine. (a) Solid cone, hollow cylinder, solid cylinder (from left to right). (b) Different size of samples.

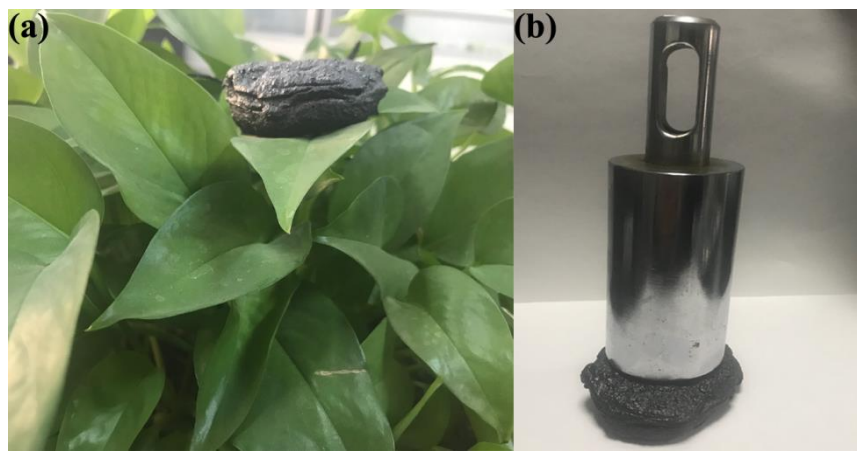

**Figure S3.** The mechanical performance test of C-FFNC sample (a) The sample is placed on the leaves of a campus plant. (b) 1 kg of weight is placed on top of the sample

Here we use the traditional carbon foam mechanical performance test method to express its mechanical properties. The carbon foam is placed on the leaves of the campus plants without obvious deformation, indicating that the quality is lighter. Furthermore, it is placed under the 1kg standard weight without any mechanical deformation or breakage, so it shows that its mechanical strength is excellent.

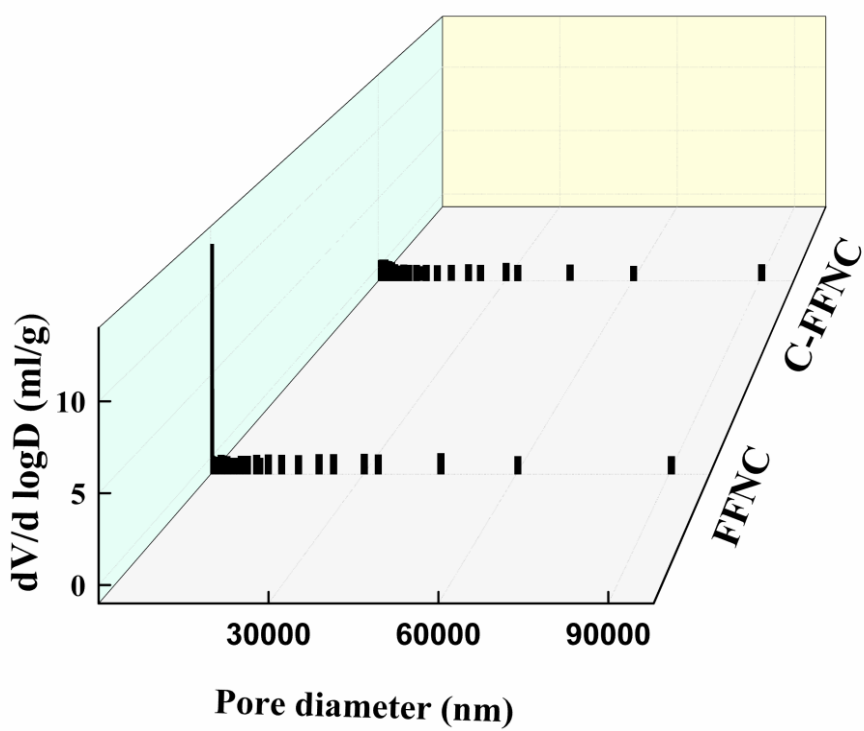

**Figure S4.** Pulverized mercury instrument raw data mapping analysis of FFNC and C-FFNC.

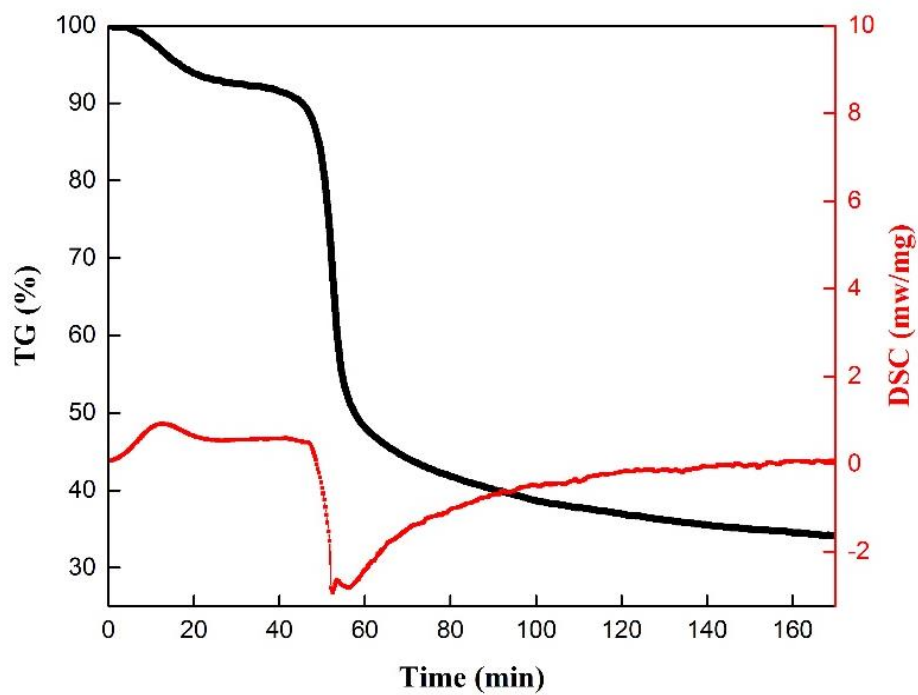

**Figure S5.** TG-DSC curve analysis of FFC samples.

Thermogravimetric (TG)-differential scanning calorimetry (DSC) analysis of freeze-dried samples can explained the substance change during the carbonization process. As shown in the figure, in the 0-60 minutes, the temperature is raised at 5 ℃ min-1, then it is kept at 300 ℃ for 60 minutes, and then naturally cooled to room temperature. The TG curve showed two processes of mass reduction, which reduced the total mass by 7.6% and 58.36% respectively. There are two corresponding peaks in the DSC curve. The first peak of DSC curve is an endothermic process with an onset temperature of 41.1 ℃ and a peak temperature of 83.6 ℃. It can therefore be inferred that the first mass reduction of the TG curve is the evaporation of water inside. The other peak of DSC curve is an exothermic process with an onset temperature of 275.1 ℃ and a peak temperature of 301.1 ℃. The main starch and protein were incompletely carbonized during this process.

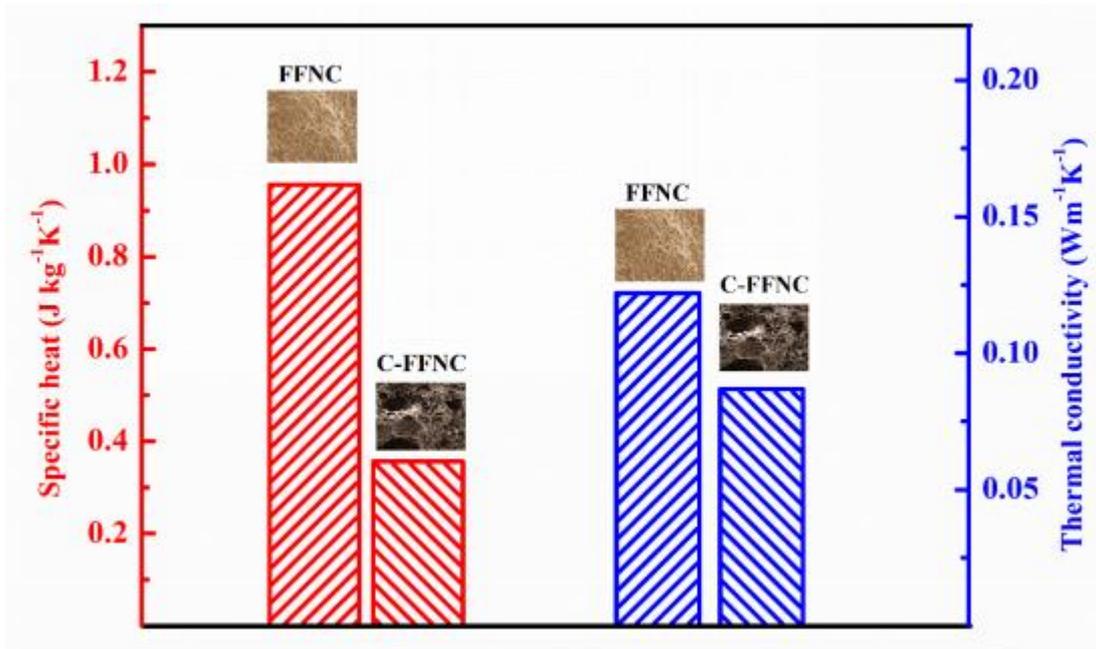

Figure S6 The specific heat and thermal conductivity of samples before and after carbonization.

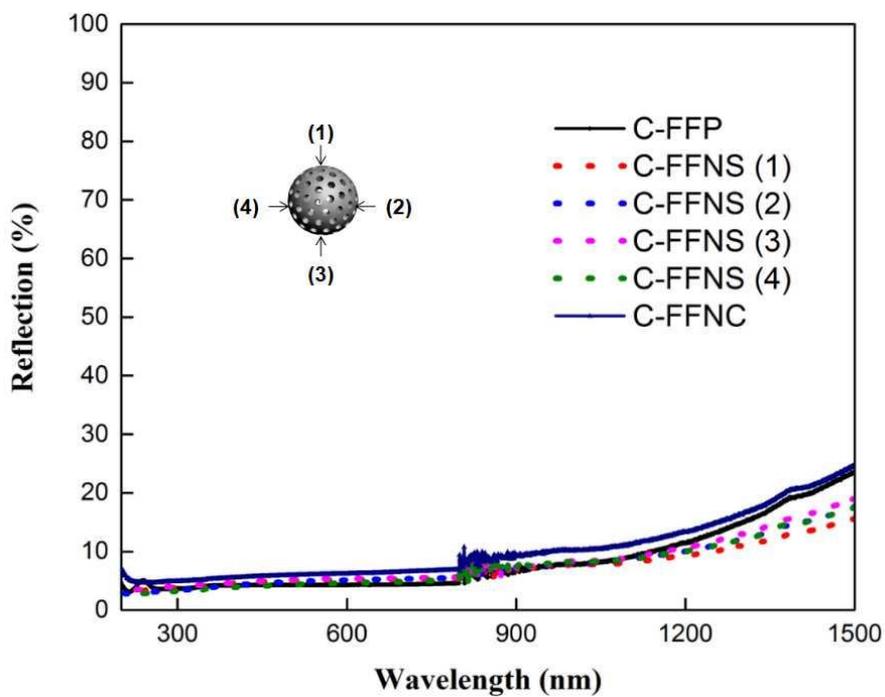

Figure S7. The reflection curve of C-FFP, C-FFNS, C-FFNC sample.

To further confirm that the optical properties of the different samples are the same, here we test the reflectance of the C-FFNC, C-FFP, and C-FFNS with four different angles, respectively. In the band (400-760 nm), there is no obvious fluctuation for all samples. However, through detailed calculations, the absorption rate of C-FFNC is slighter lower than C-FFNS and C-FFP (about 1%) due to C-FFNS and C-FFP with higher surface area.

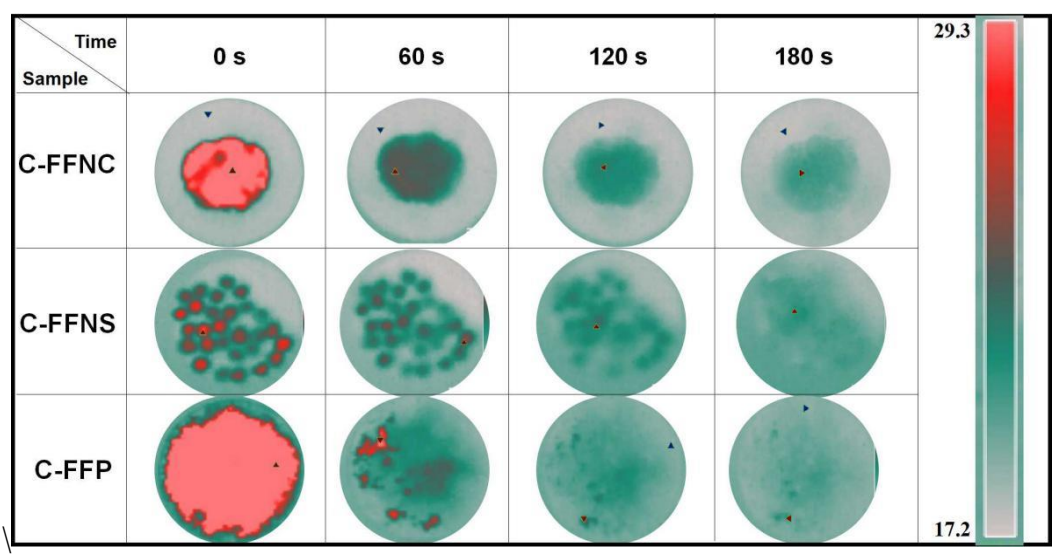

**Figure S8.** The wetting process of C-FFNC, C-FFNS, C-FFP sample at 0 s, 60 s, 120 s, 180 s.

We preheated all samples in an oven (set temperature 110 °C) for 1 min, then placed them inside the chamber and observed a temperature drop of ~4 °C in three minutes. The above heating step was repeated, and the sample was placed in water, and the surface temperature was changed within three minutes as shown in Figure S2. In the absence of sunlight, effective water transport and natural evaporation mainly promote the cooling process of the sample. The surface temperature of the three samples is close to the water temperature at three minutes. This can again prove that the transport of water is nearly close for the three samples.